\documentclass[pra,twocolumn,aps,longbibliography]{revtex4-1}
\usepackage{amsmath}
\usepackage{amssymb}
\usepackage{graphicx}
\usepackage[usenames,dvipsnames]{color}
\usepackage{braket}
\usepackage{bm}
\usepackage{times}
\usepackage{hyperref}
\usepackage{pdfpages}
\usepackage{float}
\usepackage{lipsum}
\usepackage[dvipsnames]{xcolor}
\hypersetup{
	colorlinks=true,
	citecolor=blue,
	linkcolor=blue,
	urlcolor=blue}
\usepackage{tikz}
\usetikzlibrary{decorations.markings}
\usetikzlibrary{decorations.pathmorphing,snakes}
\usetikzlibrary{arrows.meta}
\tikzset{middlearrow/.style={
	decoration={markings,
	mark= at position 0.5 with {\arrow{#1}} ,
	},
	postaction={decorate}
}
}

\makeatletter
\AtBeginDocument{\let\LS@rot\@undefined}
\makeatother

\begin{document}
%%%%%%%%%%%%%%%%%%%%%%%%%%%%%%%%%%%%%%%%%%%%%%%%%%%%%%%%%%%%%%%%%%%%%%%%%%%%%
%%%%%%%%%%%%%%            Titles and authors            %%%%%%%%%%%%%%%%%%%%%
%%%%%%%%%%%%%%%%%%%%%%%%%%%%%%%%%%%%%%%%%%%%%%%%%%%%%%%%%%%%%%%%%%%R%%%%%%%%%
\title{Disordered solid to Bose-glass transition in Bose-Hubbard model with 
       disorder and long-range interactions}
\author{Kohenjit Pebam, Laxmi Angom, Bhumika Thoudam, Momin Hrangbung,
        Deepak Gaur and Dilip Angom}
\affiliation{Manipur University,Canchipur-795003,Physics Department}
%\affiliation{Manipur University,Canchipur-795003,Physics Department}
%\affiliation{Manipur University,Canchipur-795003,Physics Department}
%\affiliation{Manipur University,Canchipur-795003,Physics Department}
%\affiliation{Manipur University,Canchipur-795003,Physics Department}
%\affiliation{Manipur University,Canchipur-795003,Physics Department}

%%%%%%%%%%%%%%%%%%%%%%%%%%%%%%%%%%%%%%%%%%%%%%%%%%%%%%%%%%%%%%%%%%%%%%%%%%%%%%
%%%%%%%%%%%%%%%                 Abstract               %%%%%%%%%%%%%%%%%%%%%%%
%%%%%%%%%%%%%%%%%%%%%%%%%%%%%%%%%%%%%%%%%%%%%%%%%%%%%%%%%%%%%%%%%%%%%%%%%%%%%%
\begin{abstract}
The introduction of disorder in Bose-Hubbard model gives rise to
new glassy quantum phases, namely the Bose-glass (BG) and disordered solid
(DS) phases. In this work, we present the rich phase diagram of interacting
bosons in disordered two-dimensional optical lattice, modelled by the
disordered Bose-Hubbard model. We systematically probe the effect of long-range
interaction truncated to the nearest neighbors and  next-nearest
neighbors on the phase diagram. We investigate the zero-temperature
ground-state quantum phases using the single-site Gutzwiller mean field
(SGMF) theory. We also employ strong-coupling perturbative expansion to
identify the nature of ground-state solid phases analytically.  At 
sufficiently high disorder strength, we observe a quantum phase transition
between the DS and BG phases. We have investigated this transition in greater
detail using cluster Gutzwiller mean field theory to study the effect of
inter-site correlations which is absent in the SGMF method. We have also 
studied this phase transition from the perspective of percolation theory.
\end{abstract}

\maketitle

%%%%%%%%%%%%%%%%%%%%%%/%%%%%%%%%%%%%%%%%%%%%%%%%%%%%%%%%%%%%%%%%%%%%%%%%%%%%%%
%%%%%%%%%%%%%%%      Introduction                %%%%%%%%%%%%%%%%%%%%%%%%%%%%
%%%%%%%%%%%%%%%%%%%%%%%%%%%%%%%%%%%%%%%%%%%%%%%%%%%%%%%%%%%%%%%%%%%%%%%%%%%%%
\section{Introduction}
The disordered systems have been a subject of several studies and continues
to attract attention. This is primarily fueled by its intimate connections 
with the real-world physical systems. These systems are
characterized by deviations from long-range ordering, which may be 
attributed to the randomness in some of the components in the system. The 
disordered quantum systems have attracted significant attention and the 
studies on these systems have led to the discovery of fascinating phenomena 
such as Anderson localization
\cite{anderson_58, evers_08} and the glassy quantum phases such as Bose-glass
\cite{fisher_89, fallani_07}, to name a few. Disordered quantum 
many-body systems of interacting bosons have been explored in various 
experiments and theoretical studies \cite{krauth_91, fallani_07, gurarie_09, 
deissler_10, pasienski_10, gao_17, lin_11, chiara_14, sukla_19}.
In the theoretical studies, these systems are modelled using the disordered
Bose-Hubbard model (DBHM) \cite{fisher_89} which is experimentally realizable
with optical lattices and speckle laser fields \cite{white_09, bouyer_10} for
creating the random disorder in the lattice potential.
In recent years, ultracold atoms in optical lattices have emerged as an ideal
platform to simulate the physics of quantum many-body systems. This is due to
the developments of novel experimental techniques to manipulate the system
parameters together with the accessibility of the strongly interacting
regime \cite{bloch_05, morsch_06, lewenstein_07, krutitsky_16, schafer_20}. 
Such a system of bosonic atoms is modelled by the Bose-Hubbard model (BHM) 
and its extensions \cite{fisher_89, jaksch_98}.

The BHM exhibits a metal-insulator transition between superfluid (SF) and Mott
insulator (MI) quantum phases which have been observed in experiments with
optical lattices \cite{greiner_02}. In the DBHM, for any strength of disorder, 
the BG phase intervenes the metal-insulator transition in the thermodynamic 
limit \cite{pollet_09}. The effects of long-range interactions can be probed 
in BHM by using dipolar atoms, molecules or Rydberg atoms in optical lattices, 
and the modified system is then referred to as the extended BHM. These 
interactions introduces a new quantum phase referred as the supersolid (SS) 
phase which have both the superfluid property as well as crystalline order 
\cite{boninsegni_12, trefzger_08, rossini_12, yamamoto_12, suthar_20}. 
This novel phase has been recently observed in the experiments
\cite{morales_17, tanzi_19, bottcher_19, chomaz_19}. The introduction of
disorder leads to emergence of other novel quantum phases called disordered 
solid and BG phase. In this work, we report the rich phase 
diagram of the disordered extended Bose-Hubbard model (DEBHM). We have studied
the effect of dipolar interactions, beyond the standard truncation to nearest 
neighbors (NN) only \cite{gao_17}, and have studied the effect of next-nearest
neighbors (NNN) on the phase diagram. This gives a richer phase diagram with 
insulating phases having average occupancies $\rho =1/4$ and $\rho =3/4$, 
which is absent in the NN truncation.

In the present work, we employ the SGMF theory
\cite{rokhsar_91,sheshadri_93} to study the ground-state quantum phases of
softcore dipolar bosons in $2D$ square optical lattices. The DEBHM exhibits a
rich phase diagram comprising of six types of quantum phases, namely the MI,
density wave (DW), DS, BG, SS and SF phases at zero temperature and are
demarcated by various quantum phase transitions. Using strong coupling
perturbative expansion, we also examine the nature of the insulating ground 
state among the checkerboard and stripe geometry for density wave phases and 
the results agree with the numerical results obtained using SGMF method. 
Earlier works had reported the transition from DS to BG in the hardcore 
limit\cite{zhang_jan20}. Our results using SGMF method, shows this phase 
transition for the case of softcore bosons, which we study in better detail by 
accounting the effect of quantum correlations using the cluster Gutzwiller 
mean-field (CGMF) theory\cite{luhmann_13}. The compressibility and structure 
factor calculations are also done using CGMF and comparison is made with the 
results obtained using SGMF. The CGMF results suggests an
enlargement of the BG phase domain whereas domain of DS phase shrinks. 
The other approach to characterize the DS and BG phases is to focus on the 
background phase. In short, DS phase has a structured background whereas it
is uniform in BG phase. Considering this we analyze the DS-BG phase transition
from the viewpoint of the percolation transition and investigate the critical 
value of chemical potential when the spanning cluster emerges in the systems. 
It is to be mentioned that the concepts and tools from percolation theory 
have wide applicability across various types of systems
\cite{broadbent_57, flory_53, elliott_60, stauffer_92}.
The paper is organized as follows. In Sec. II, we introduce the model
Hamiltonian and provide a description of the SGMF method used to obtain
the zero temperature ground-state phase diagrams in our study. The various 
order parameters employed to differentiate the quantum phases from each 
other are also described here. Sec. III presents the strong-coupling 
perturbative expansion results involving comparison of the density 
wave phases to identify the structure of the ground-state solid quantum 
phases. In Sec. IV, we present the ground state phase diagrams of DEBHM at 
different disorder strengths, for the cases involving truncation of dipolar 
interactions till NN and NNN sites. Here, we also discuss on the DS-BG phase 
transition. Finally, Sec. V  presents the conclusion of our work.

%%%%%%%%%%%%%%%%%%%%%%%%%%%%%%%%%%%%%%%%%%%%%%%%%%%%%%%%%%%%%%%%%%%%%%%%%%
%%%%%%%%%%%%%      Section: Theoretical methods   %%%%%%%%%%%%%%%%%%%%%%%%
%%%%%%%%%%%%%%%%%%%%%%%%%%%%%%%%%%%%%%%%%%%%%%%%%%%%%%%%%%%%%%%%%%%%%%%%%%
\section{Theoretical methods}
A system of ultracold bosons with long-range interaction and disorder
in a square optical lattice is well described by the DEBHM. 
The Hamiltonian of the system is 
\begin{eqnarray}
	\hat{H}_{\text{DEBHM}}&= &-\sum_{p,q}\left(J_{x}\hat{b}^{\dagger}_{p+1,q}
	\hat{b}_{p,q}
	+J_{y}\hat{b}^{\dagger}_{p,q+1}
	\hat{b}_{p,q}+\text{H.c.}\right)
	\nonumber \\ 
	&  &+\sum_{p,q}\hat{n}_{p,q}
	\left[\frac{U}{2}\left(\hat{n}_{p,q}-1\right)
	-\tilde{\mu}_{p,q}\right]                                     
	\nonumber\\                     
	&  &+\sum_{\xi,\xi'}\frac{V_{\xi,\xi'}}{2}
	\hat{n}_{\xi}\hat{n}_{\xi'},
	\label{ham}
\end{eqnarray} 
where $p(q)$ is the lattice site index along the $x(y)$ direction,             
$\hat{b}_{p,q}^{\dagger}(\hat{b}_{p,q})$ is the boson creation
(annihilation) operator at the lattice site $(p,q)$, $\hat{n}_{p,q}
= \hat{b}^{\dagger}_{p,q}\hat{b}_{p,q}$  is the number operator for bosons, 
$J_{x}$  and $J_{y}$ are the tunneling strength along the $x$ and $y$ 
directions respectively. The effective chemical potential 
$\tilde{\mu}_{p,q}=\mu-\epsilon_{p,q}$ consist of the average chemical potential 
$\mu$ and the  disorder $\epsilon_{p,q}$. The disorder is a random energy 
offset introduced using random numbers  distributed 
uniformly with $\epsilon_{p,q}\in[-D,D]$ and $D$ is the bound of random numbers. 
The on-site interaction energy is $U>0$. For compact notations, 
$\xi\equiv(p,q)$ and $\xi'\equiv(p',q')$ represent the lattice 
indices of the neighbouring sites. 
To represent the long range potential we consider the $V_1-V_2$ model 
\cite{yamamoto_12}, which has the expression
\begin{equation}
	V_{\xi,\xi'}=
	\begin{cases}
		V_{1} & \text{if}\, |\textbf{r}_{\xi}-\textbf{r}_{\xi'}|=a, \\
		V_{2} & \text{if}\, |\textbf{r}_{\xi}-\textbf{r}_{\xi'}|=
		\sqrt{2}a,\\
		0     & \text{otherwise},
	\end{cases}
   \label{v1_v2_model}
\end{equation}
where $a$ is the lattice constant and $\textbf{r}_{\xi}$ and 
$\textbf{r}_{\xi'}$ are the position vectors of the $\xi$ and $\xi'$
lattice sites, respectively. $V_{1}\geqslant0$ and $V_{2}\geqslant0$ are the 
nearest neighbour (NN) and next nearest neighbour (NNN) interactions. 
Here, $V_{1}$ and $V_{2}$ are related through the relation 
$V_{2}/V_{1}=1/(2\sqrt{2})$. This relation corresponds to the inverse cube 
power law of the isotropic dipole-dipole interaction.

%%%%%%%%%%%%%%%%%%%%%%%%%%%%%%%%%%%%%%%%%%%%%%%%%%%%%%%%%%%%%%%%%%%%%%%%%%%%%%
%%%%% Subsection:Single-site Gutzwillwer mean-field (SGMF) theory %%%%%%%%%%%%
%%%%%%%%%%%%%%%%%%%%%%%%%%%%%%%%%%%%%%%%%%%%%%%%%%%%%%%%%%%%%%%%%%%%%%%%%%%%%%

\subsection{Single-site Gutzwillwer mean-field (SGMF) theory}
The SGMF theory is a mean-field theory 
employed to study the ground state of the lattices like the case of disordered 
extended Bose-Hubbard model which is described by Eq.(\ref{ham}). In the SGMF 
theory, the bosonic operators are approximated as sum of their expectation 
values and quantum fluctuations
\begin{eqnarray}
	\hat{b}_{\xi}          &=&\phi_{\xi}+\delta \hat{b}_{\xi}, 
	\nonumber \\
	\hat{b}^{\dagger}_{\xi}&=&\phi^{*}_{\xi}+\delta\hat{b}^{\dagger}_{\xi}.
\end{eqnarray}
Then the product of the creation and the annihilation operators
can be written as
\begin{equation}
	\hat{b}^{\dagger}_{\xi}\hat{b}_{\xi'}
	\approx \phi^{*}_{\xi}\hat{b}_{\xi'}
	+\hat{b}^{\dagger}_{\xi}\phi_{\xi'}
	-\phi^{*}_{\xi}\phi_{\xi'}.
\end{equation}
Here, $\phi_{\xi}=\langle\hat{b}_{\xi}\rangle$ is the SF order
parameter at the lattice site $\xi$. Similarly, like earlier, to find the mean 
field approximation of the long range part, we expand $\hat{n}_{\xi}$ and
$\hat{n}_{\xi'}$ as the sum of their expectation values and quantum 
fluctuations
\begin{eqnarray}
	\hat{n}_{\xi}&=&\langle \hat{n}_{\xi}\rangle    +\delta\hat{n}_{\xi},
	\nonumber \\
	\hat{n}_{\xi'}&=&\langle \hat{n}_{\xi'}\rangle  +\delta\hat{n}_{\xi'}.
	\nonumber
\end{eqnarray}
Neglecting the fluctuations, the density dependence in the long-range 
interaction can be written as
\begin{equation}
	\hat{n}_{\xi}\hat{n}_{\xi'}=\hat{n}_{\xi}\langle \hat{n}_{\xi'}\rangle
	 \hat{n}_{\xi'}\langle \hat{n}_{\xi}\rangle
	-\langle \hat{n}_{\xi}\rangle
	\langle \hat{n}_{\xi'}\rangle.
\end{equation}
Using the above approximations, the single-site mean-field Hamiltonian 
of the system is
\begin{eqnarray}
	\hat{H}^{\text{MF}}_{p,q} &=&-\left[J_{x}\left(\phi^{*}_{p+1,q}
	\hat{b}_{p,q}
	-\phi^{*}_{p+1,q}\phi_{p,q}\right)
	\right.
	\nonumber\\
	& &\left. +J_{y}\left(\phi^{*}_{p,q+1}
	\hat{b}_{p,q}
	-\phi^{*}_{p,q+1}\phi_{p,q}\right)
	+\text{H.c.}\right]\nonumber \\
	& &+\hat{n}_{p,q}
	\left[\frac{U}{2}
	\left(\hat{n}_{p,q}-1\right)
	-\tilde{\mu}_{p,q}
	\right]\nonumber\\                     
	& &+\sum_{\xi'}V_{\xi,\xi'}
	\left(\hat{n}_{\xi}\langle\hat{n}_{\xi'}
	\rangle-\langle\hat{n}_{\xi}\rangle
	\langle\hat{n}_{\xi'}\rangle\right),
\end{eqnarray}
The mean-field Hamiltonian of the system can be written as the sum of the 
single-site mean-field hamiltonians
\begin{equation}
	\hat{H}_{\text{MF}}=\sum_{p,q}\hat{H}^{\text{MF}}_{p,q}.
\end{equation}
With this approximation, the neighbouring lattice sites are coupled through 
the SF order parameter and the eigenstate of a single-site is a linear 
combination of the Fock states. Accordingly, the eigenstate of the whole 
system is the direct product of the single-site states. Thus, with the 
Gutzwiller ansatz, the many-body wave function of the ground state is 
\begin{equation}
	\ket{\Psi} =\prod_{p,q}\ket{\psi}_{p,q}
	=\prod_{p,q}\left(\sum_{n=0}^{N_{b}-1}
	c^{(p,q)}_{n}\ket{n}_{p,q}\right),
\end{equation}
where $\ket{\psi}_{p,q}$ is the single site ground state, $N_{b}$ is
the dimension of the Fock space, $\ket{n}_{p,q}$ is the Fock state of
$n$ bosons occupying site $(p,q)$, and $c^{(p,q)}_{n}$ are the coefficients of
the linear combination. The normalization condition of the wave function is
$\sum_{n}|c^{(p,q)}_{n}|^{2}=1$. The SF order parameter at a lattice site
can be calculated as
\begin{equation}
	\phi_{p,q}=\bra{\Psi}\hat{b}_{p,q}\ket{\Psi}
	=\sum_{n=0}^{N_{b}-1}\sqrt{n}c^{(p,q)*}_{n-1}c^{(p,q)}_{n}.
\end{equation}
However, as a measure of global system properties we use the system size
averaged $\phi$ which is defined as
\begin{equation}
	\phi=\frac{\sum_{p,q}\left|\phi_{p,q}\right|}{N},
\end{equation}
where $N$ is the total number of lattice sites in the system. Similarly,
we define the disordered averaged SF order parameter 
\begin{equation}
	\Phi=\frac{1}{N_{d}}\sum_{i=1}^{N_{d}}\phi_i,
\end{equation}
where, $N_{d}$ is the number of disorder samples and $\phi_i$ is the system 
size averaged SF order parameter of the $i$th sample. Similarly, the 
lattice site occupancy is 
\begin{equation}
	n_{p,q}=\bra{\Psi}\hat{b}^{\dagger}_{p,q}\hat{b}_{p,q} \ket{\Psi}
	=\sum_{n=0}^{N_{b}-1}n|c^{(p,q)}_{n}|^{2}.
\end{equation}
The quantities $\phi$ and $n_{p,q}$ are used to identify the quantum
phases of the system. Our previous works \cite{bandyopadhyay_19,bai_20} may
be referred for additional details on the single-site Gutzwiller 
mean-field method we have used in the present study.

%%%%%%%%%%%%%%%%%%%%%%%%%%%%%%%%%%%%%%%%%%%%%%%%%%%%%%%%%%%%%%%%%%%%%%%%%%
%%%%%%%%%%%% subsection:Characterization of phases %%%%%%%%%%%%%%%%%%%%%%%
%%%%%%%%%%%%%%%%%%%%%%%%%%%%%%%%%%%%%%%%%%%%%%%%%%%%%%%%%%%%%%%%%%%%%%%%%%

\subsection{Characterization of phases}
The ground state of DEBHM can be of different quantum phases depending
on the parameters of the system. And, the quantum phases can be distinguished 
using order parameters which are measure of different properties. In the
present work, we use superfluid fraction $f_{s}$, $\phi$, 
%lattice site occupancy%, superfluid stiffness $\rho_{s}$, 
%relative average occupancy $\langle \Delta n\rangle$, 
structure factor 
$S(\vec{q})$ and compressibility $\kappa$. For a quantum phase which is phase 
coherent like the SF, a finite energy is required to destroy the phase 
coherence. This implies that the SF phase acquires stiffness towards phase 
change or has phase rigidity. For the calculation of $f_{s}$, a twisted 
boundary condition(TBC) is imposed on the state such that when it is applied 
along $x$ direction, the hopping term at the boundary is transformed as
\begin{eqnarray}
	J_{x}\left(\hat{b}^{\dagger}_{p+1,q}\hat{b}_{p,q}\right)
	\rightarrow J_{x}\left(\hat{b}^{\dagger}_{p+1,q}\hat{b}_{p,q}
	e^{i2\pi\varphi}\right),
\end{eqnarray}
where $\varphi$ is the twist applied to hopping at the boundary the 
periodic boundary condition, $L$
is the lattice size along $x$ direction and $i2\pi\varphi$ is the phase shift
possesed by an atom when it tunnels between the nearest neighbour lattice
sites. With the introduction of TBC, the DEBHM Hamiltonian is modified and 
let the corresponding groundstate energy be $E_{\varphi}$. Let $E_{0}$ be the
energy eigenvalue corresponding to DEBHM Hamiltonian without TBC. Then
superfluid fraction $f_{s}$, expressed in terms of the DEBHM Hamiltonian 
parameters, is given as \cite{roth_03}
\begin{equation}
	f_{s}=\frac{I^{2}}{NJ\varphi^{2}}[E_{\varphi}-E_{0}],
	\label{sf-stiffness}
\end{equation}
where $I$ is the total lattice sites along a direction and $N$ is the total
occupancy of the system. Among the ground state quantum phases, MI and DW 
are incompressible and the remaining quantum phases DS, BG, SS and SF are 
compressible.  The compressibility of the system is defined  using the 
following relation
\begin{equation}
	\kappa=\frac{\partial \langle\hat{n}\rangle}{\partial \mu}.
\end{equation}
%The relative average occupancy $\langle\Delta n\rangle$ is another order 
%parameter which can distinguish the phases with diagonal long-range order
%DW and SS from the uniform phases MI and the SF. In the SGMF method,
%for a $K\times L$ lattice, it is defined as
%\begin{equation}
%	\langle\Delta n\rangle=\frac{1}{K\times L}\sum_{\langle\xi,\xi'\rangle}
%	|\langle\hat{n}_{\xi}\rangle-\langle\hat{n}_{\xi'}\rangle|.
%	\label{rel_occupancy}
%\end{equation}
The occupancies in the DW phase is a combination of two sublattices A and
B with different fillings. This description implies that in the above 
equation $\langle\hat{n}_{\xi}\rangle \equiv \langle\hat{n}_{A}\rangle$
and $\langle\hat{n}_{\xi'}\rangle \equiv \langle\hat{n}_{B}\rangle$
are the sublattice occupancies of the two sublattices. Another order 
parameter which we used to determine the diagonal long-range order, 
characteristic of a solid or structured phase, is the structure factor 
$S(\vec{q})$. It is defined in terms of the density-density correlation 
function
\begin{equation}
	S(\vec{q})=\frac{1}{N}\sum_{\xi,\xi'}e^{i\vec{q}.\left(\vec{r}_{\xi}
	-\vec{r}_{\xi'}\right)}\langle \hat{n}_{\xi}\hat{n}_{\xi'}\rangle.
\end{equation}
As an example consider the checkerboard density wave phase. One of it's 
distinguishing property is a peak of $S(\vec{q})$ at the wave vector 
$\vec{q}=(\pi,\pi)$. The order parameters employed to distinguish 
various quantum phases exhibited 
by DEBHM are listed in 
Table~\ref{order_par_tab}.
\begin{table}[h]
  %  \begin{center}
  \setlength{\tabcolsep}{14pt}
  \renewcommand{\arraystretch}{2}
  \caption{Table showing values of different order parameters
  of different quantum phases exhibited by DEBHM.}
  \begin{tabular}{cccc}
     \hline \hline
     Quantum phases&$\left|f_{s}\right|$&$ \left|S(\pi,\pi)\right| $&$
     \phi$                \\
     \hline \hline
     $\frac{1}{4}$DW & 0 & $\neq 0$ (integer) &0 \\
     % \hline
     CDW & 0 & $\neq 0$ (integer) &0 \\
     % \hline
     $\frac{3}{4}$DW &0&$\neq 0$ (real) & 0\\
     % \hline
     SS &$\neq 0$&$\neq 0$ (real)&$\neq 0$\\
     % \hline
     SF&$\neq 0$ &=0&$\neq 0$\\
     % \hline
     BG&=0 &=0&$\neq 0$\\ 
     % \hline
     DS&=0 &$\neq 0$ (real)&$\neq 0$\\
     % \hline
     MI&=0&=0&=0\\ 
     \hline 
  \end{tabular}
 \label{order_par_tab}
  % \caption{Table showing values of different order parameters
  %  of different quantum phases exhibited by DEBHM.}
  % \end{center}
\end{table}

%%%%%%%%%%%%%%%%%%%%%%%%%%%%%%%%%%%%%%%%%%%%%%%%%%%%%%%%%%%%%%%%%%%%%%%%%%%%%%
%%%%%%%%%% Strong-Coupling perturbative expansion %%%%%%%%%%%%%%%%%%%%%%%%%%%%
%%%%%%%%%%%%%%%%%%%%%%%%%%%%%%%%%%%%%%%%%%%%%%%%%%%%%%%%%%%%%%%%%%%%%%%%%%%%%%

\section{Strong-coupling perturbative expansion}
To identify the  quantum phase of the ground state analytically we use
strong-coupling perturbation theory and apply it to the extended 
Bose-Hubbard model with both NN and NNN long-range interactions
\cite{iskin_09}. In the perturbative calculations we consider 
the insulating phases CDW and SDW 
unperturbed state. To identify the quantum phases of the system, we consider 
the system size as a 4$\times$4 square lattice. In this approach, the 
tunneling term in the extended Bose-Hubbard Hamiltonian is considered as the 
perturbation and the remaining terms as the unperturbed Hamiltonian
\cite{rashi_12,iskin_09,freericks_96}. We, then, 
apply the many-body version of time-independent perturbation theory and expand 
the energy of the insulating phases in the powers of the tunneling parameter 
$J$. Due to the constraints on the choice of the unperturbed state, the 
method cannot be used to determine the phase boundary between two 
compressible phases. Using the method the perturbative expansion is calculated
upto second order\cite{wang_18} 
for the CDW without extra particle as well as for SDW without
extra particle. The extended Bose-Hubbard Hamiltonian is given by
\begin{eqnarray}
	\hat{H}_{\text{eBHM}} &=&-J\sum_{\langle \xi,\xi'\rangle}
	\left(\hat{b}^{\dagger}_{\xi}\hat{b}_{\xi'}
	+\hat{b}^{\dagger}_{\xi'}\hat{b}_{\xi}\right)
	+\sum_{\xi}\bigg[\frac{U}{2}\hat{n}_{\xi}
	(\hat{n}_{\xi}-1)
	                   \nonumber\\
	& &-\mu\hat{n}_{\xi}\bigg]
	+\sum_{\xi,\xi'}\frac{V_{\xi \xi'}}{2}
	\hat{n}_{\xi}\hat{n}_{\xi'} \nonumber\\
	&=&\hat{H'}+\hat{H_{0}}.
\end{eqnarray}
In the Hamiltonian given above, $H'=-J\sum_{\langle \xi,\xi'\rangle}
(\hat{b}^{\dagger}_{\xi}\hat{b}_{\xi'}
+\hat{b}^{\dagger}_{\xi'}\hat{b}_{\xi})$ is the perturbation term of the
Hamiltonian and the remaining terms are the unperturbed terms $\hat{H}_{0}$
\cite{iskin_09,rashi_12}.

%%%%%%%%%%%%%%%%%%%%%%%%%%%%%%%%%%%%%%%%%%%%%%%%%%%%%%%%%%%%%%%%%%%%%%%%%%%%%%
%%%%   subsection:Ground-state wave functions of the insulating phases    %%%%
%%%%%%%%%%%%%%%%%%%%%%%%%%%%%%%%%%%%%%%%%%%%%%%%%%%%%%%%%%%%%%%%%%%%%%%%%%%%%%

\subsection{Ground-state wave functions of the insulating phases}
 With long-range interactions, DW phase can emerge as the ground
state in certain parameter regimes since it lowers the repulsion between
neighboring lattice sites. For the $V_1 - V_2$ \cite{yamamoto_12} 
model given by Eq.~(\ref{v1_v2_model}) with $V_1 > 0$ and $V_2 = 0$, 
the checkerboard ordering
lowers the repulsion between NN sites, resulting in the CDW phase as the 
ground state. Considering $V_1 = 0$ and $V_2 > 0$, the stripe ordering is 
preferred as it lowers the repulsion between NNN sites can result in a lower 
energy. For intermediate values of $V_1$ and $V_2$, the CDW and SDW phases 
needs to be compared for lower energy. In the SGMF theory, the generic
form of the ground-state wave functions of the CDW and SDW phases is \cite{iskin_09}
\begin{equation}
	\ket{\Psi^{(0)}_{\text{DW}}} =\prod_{\xi\in A, \xi' \in B}
	\ket{n_{A}}_{\xi} \ket{n_{B}}_{\xi'},
\end{equation}
where, $n_{A}$ and $n_{B}$ as mentioned earlier are the occupancies at 
the lattice sites of sublattice $A$ and $B$, respectively. More specifically,
for the case of CDW phase the site indices $\xi$ and $\xi'$ satisfy the 
following conditions
\begin{eqnarray}
 \xi  &&\equiv (p, q)  : p +q  \text{ is even}, \nonumber \\
 \xi' &&\equiv (p',q') : p'+q' \text{ is odd}. 
\end{eqnarray}
while for SDW phase similar relations, with stripes running along the $y$-axis,
are 
\begin{eqnarray}
 \xi  &&\equiv (p, q)  : p  \text{ is even}, \nonumber \\
 \xi' &&\equiv (p',q') : p' \text{ is odd}. 
\end{eqnarray}
Other structured solid phases emerge when other terms in the long-range
interactions are considered \cite{trefzger_08}. However, considering NN and
NNN interactions we can probe the key features associated with combined 
effects of long-range interactions and disorder. 
\begin{figure}[h]
   \begin{center}
      \includegraphics[width=8.5 cm]{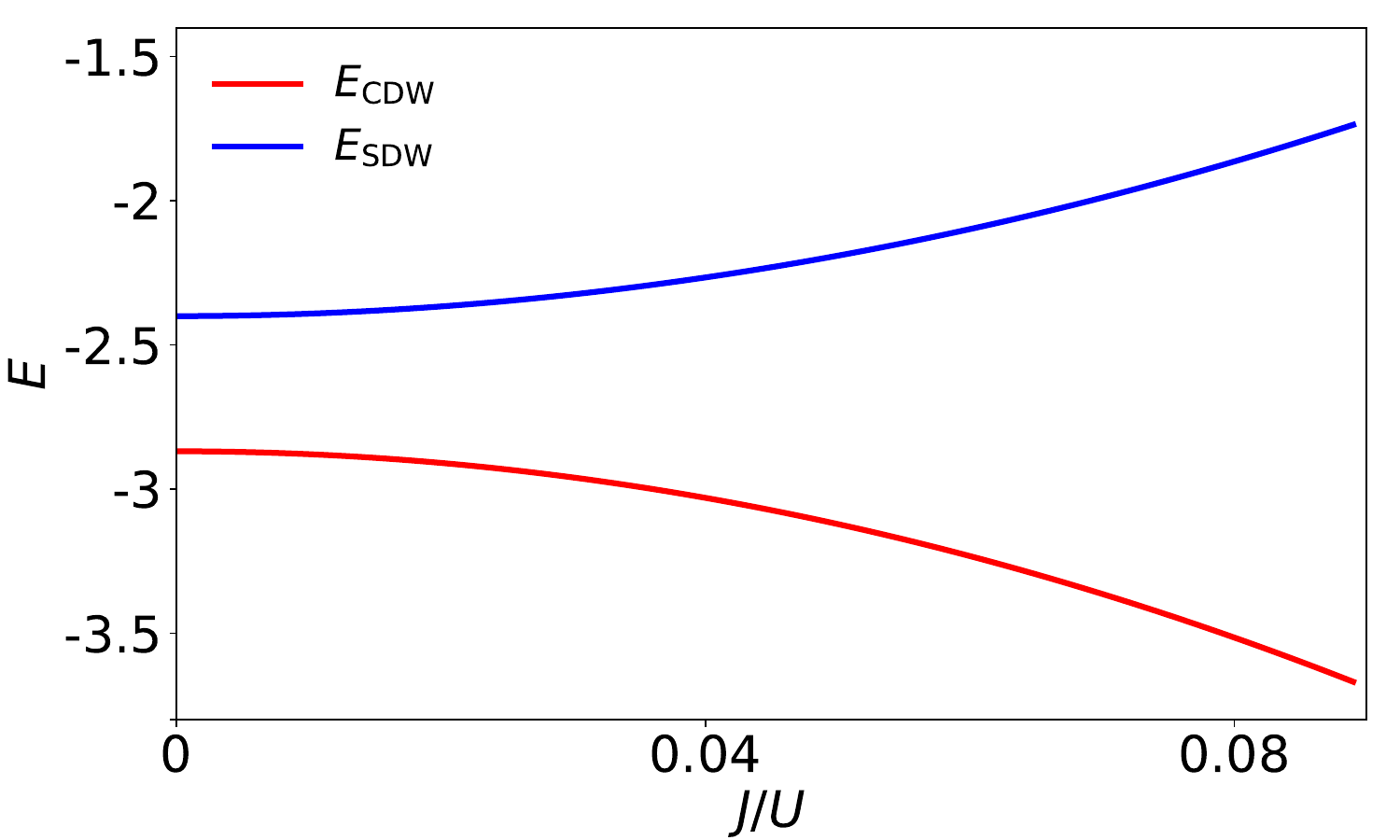}
      \caption{Energy versus $J$ plot showing comparison of $E_{\text{CDW}}$ 
               and $E_{\text{SDW}}$ at $\mu=0.5$, $V_{1}/U=0.2$ and 
               $V_{2}/U=0.0707$. The blue curve is $E_{\text{SDW}}$ and the 
               red curve is for $E_{\text{CDW}}$. Here, $E_{\text{CDW}}$ is 
               found to be of lower energy compared to $E_{\text{SDW}}$.}
      \label{ener_compare}
   \end{center}
\end{figure}

%%%%%%%%%%%%%%%%%%%%%%%%%%%%%%%%%%%%%%%%%%%%%%%%%%%%%%%%%%%%%%%%%%%%%%%%%%%%%%
%%%%%%%%%%  Ground-state energies upto second order in $J$ %%%%%%%%%%%%%%%%%%%
%%%%%%%%%%%%%%%%%%%%%%%%%%%%%%%%%%%%%%%%%%%%%%%%%%%%%%%%%%%%%%%%%%%%%%%%%%%%%%
\subsection{Ground-state energies upto second order in $J$}
In the strong-coupling perturbation expansion, we use the many-body version 
of the Rayleigh-Schrodinger perturbation theory. The perturbative expansion, 
as mentioned earlier, is in orders of $J$ with respect to the ground state of 
the system. From the perturbation theory, the first order correction to the 
ground state energy is \cite{iskin_09}
\begin{equation}
	E^{(1)}_{\text{DW}}=\bra{\Psi^{(0)}_{\text{DW}}}\hat{H}_{0}
	\ket{\Psi^{(0)}_{\text{DW}}}.
\end{equation}
Similarly, the second order correction to the ground state energy is given 
by the expression\cite{wang_18}
\begin{equation}
	E^{(2)}_{\rm DW}=\sum_{\Psi^{\text{(ex)}} \neq \Psi^{(0)}_{\text{DW}}}
	\frac{\bra{\Psi^{(0)}_{\text DW}}\hat{H'}
	\ket{\Psi^{\text{(ex)}}} \bra{\Psi^{\text{(ex)}}} \hat{H'}\ket{\Psi^{(0)}_{\text DW}}}
	{(E^{(0)}_{\rm DW}-E)},
\end{equation}
where $\ket{\Psi^{\text {(ex)}}}$ are excited states occuring as an intermediate state
in the perturbation expansion. These are also the resulting states from
the perturbation $\hat{H'}$ operating on the ground state 
$\ket{\Psi^{(0)}_{\rm DW}}$ and $E$ is the energy of the intermediate 
excited state $\ket{\Psi^{(\text{ex})}}$. We can, thus, define an intermediate excited 
state as 
\begin{equation}
	\ket{\Psi^{\text{(ex)}}}=\frac{\hat{b}^{\dagger}_{\xi}\hat{b}_{\xi'}}
	{\sqrt{(n_\xi +1)n_{\xi'}}}
	\ket{\Psi^{(0)}_{\text{DW}}},
\end{equation}
where, $\xi$ and $\xi'$ are the nearest neighbour
lattice sites. Depending on the occupancies arising from different nearest-
neighbour hopping, the CDW and SDW phases have two and four types of excited 
states, respectively. Thus, the intermediate excited states arising from the 
CDW phase are
\begin{eqnarray}
	\ket{\Psi^{\text{(ex)}}}_{A^-B^+}
	&=& \ket{n_{A}-1}_{\eta} \ket{n_{B}+1}_{\eta'}
	\prod_{\substack{\xi\in A,\xi'\in B\\
	\xi\neq \eta,\xi'\neq \eta'}}
	\ket{n_{A}}_{\xi}\ket{n_{B}}_{\xi'},
	\nonumber \\
	\ket{\Psi^{\text{(ex)}}}_{A^+B^-}
	&=& \ket{n_{A}+1}_{\eta} \ket{n_{B}-1}_{\eta'}
	\prod_{\substack{\xi\in A,\xi'\in B\\
	\xi\neq \eta,\xi'\neq \eta'}}
	\ket{n_{A}}_{\xi}\ket{n_{B}}_{\xi'}.
	\nonumber \\
\end{eqnarray}
Here, the subscripts $A^+ (A^-)$ denotes state with an addition (removal) 
of a boson at a site belonging to sublattice A. Similarly, the intermediate 
excited states of the SDW phase are 
\begin{eqnarray}
	\ket{\Psi^{\text{(ex)}}}_{A^+B^-}
	&=& \ket{n_{A}+1}_{\eta}\ket{n_{B}-1}_{\eta'}
	\prod_{\substack{\xi \in A, \xi'\in B\\ 
	\xi\neq \eta, \xi'\neq \eta'}}
	\ket{n_{A}}_{\xi}\ket{n_{B}}_{\xi'},
	\nonumber \\
	\ket{\Psi^{\text{(ex)}}}_{B^-B^+}
	&=& \ket{n_{B}-1}_{\eta}\ket{n_{B}+1}_{\eta'}
	\prod_{\substack{\xi \in A, \xi'\in B\\ 
	\xi'\neq \eta,\eta'}}
	\ket{n_{A}}_{\xi}\ket{n_{B}}_{\xi'},
	\nonumber \\
	\ket{\Psi^{\text{(ex)}}}_{A^-B^+}
	&=& \ket{n_{A}-1}_{\eta}\ket{n_{B}+1}_{\eta'}
	\prod_{\substack{\xi \in A, \xi'\in B\\ 
	\xi\neq \eta,\xi' \neq \eta'}}
	\ket{n_{A}}_{\xi}\ket{n_{B}}_{\xi'},
	\nonumber \\
	\ket{\Psi^{\text{(ex)}}}_{A^-A^+}
	&=& \ket{n_{A}-1}_{\eta}\ket{n_{A}+1}_{\eta'}
	\prod_{\substack{\xi \in A, \xi'\in B\\ 
	\xi\neq \eta,\eta'}}
	\ket{n_{A}}_{\xi}\ket{n_{B}}_{\xi'}.
	\nonumber \\
\end{eqnarray}   
In the above expressions, $\eta$ and $\eta'$ are the lattice sites where
addition and removal of boson occurs respectively. The energies of CDW phase 
and SDW phase are calculated using the non-degenerate perturbation theory. 
These are given as
\begin{eqnarray}
	E_{\text{CDW}}&=&\Big[\frac{8}{2}n_{A}(n_{A}-1)U+\frac{8}{2}n_{B}(n_{B}-1)U
	-8\mu(n_{A}+n_{B}) \nonumber\\
	&&  +32n_{A}n_{B}V_{1}
	+16n_{A}^{2}V_{2}+16n_{B}^{2}V_{2}\Big]\nonumber\\
	& & +\frac{32n_{A}(n_{B}+1)J^{2}}
	{(n_{A}-n_{B}-1)U+ V_{1}
	+4(n_{A}-n_{B})(V_{2}-V_{1})}\nonumber\\
	& & +\frac{32n_{B}(n_{A}+1)J^{2}}  
	{(n_{B}-n_{A}-1)U+ V_{1}
	+4(n_{B}-n_{A})(V_{2}-V_{1})}\nonumber 
	\label{energy_checkerboard}
	\\~\\
	E_{\text{SDW}} &=&\Big[\frac{8}{2}n_{A}(n_{A}-1)U
	+\frac{8}{2}n_{B}(n_{B}-1)U-8\mu(n_{A}+n_{B})
	\nonumber\\
	&& +8n_{A}^{2}V_{1}+8n_{B}^{2}V_{1}
	+16n_{A}n_{B}V_{1}+32n_{A}n_{B}V_{2}
	\Big]\nonumber \\
	& &+\frac{16n_{B}(n_{A}+1)J^{2}}
	{(n_{B}-n_{A}-1)U+V_{1}+(n_{A}-n_{B})V_{2}} \nonumber \\
	&&   +\frac{16n_{A}(n_{B}+1)J^{2}}{(n_{A}-n_{B}-1)U+V_{1}
	+(n_{B}-n_{A})V_{2}} \nonumber \\
	&&   +\frac{16n_{B}(n_{B}+1)J^{2} 
	+ 16n_{A}(n_{A}+1)J^{2}}{-U+V_{1}}.
	\label{energy_striped}
\end{eqnarray}
\begin{figure}[h] 
    \includegraphics[width=8.5cm]{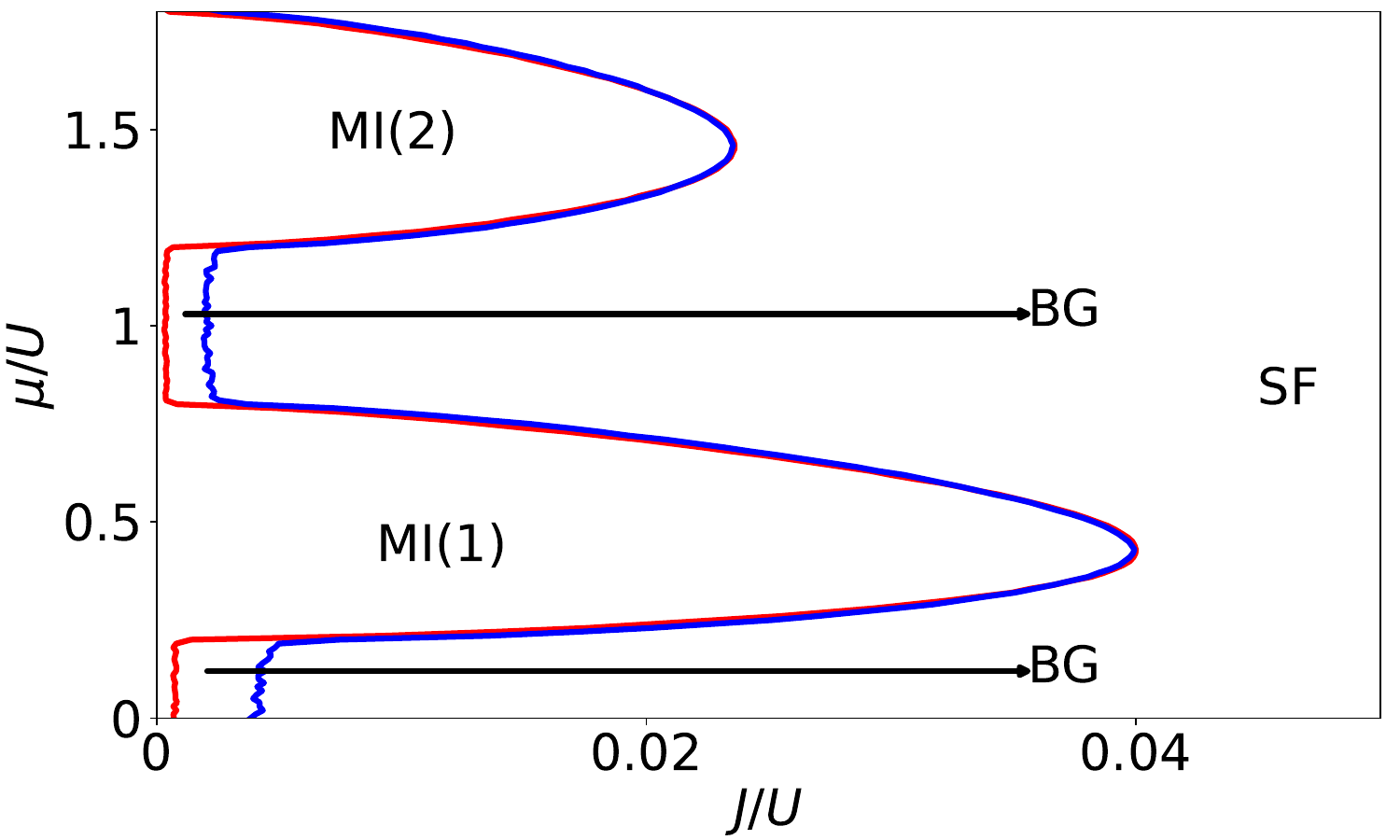}
    \caption{Ground state phase diagram of DBHM
         at disorder strength $D/U=0.2$.The red and blue 
         phase boundaries indicates the MI-BG and BG-SF phase
         boundaries respectively.}
    \label{phase-dbhm}
\end{figure}

%%%%%%%%%%%%%%%%%%%%%%%%%%%%%%%%%%%%%%%%%%%%%%%%%%%%%%%%%%%%%%%%%%%%%%%%%%%%%%
%%%%%%%%%%          Results and discussions                %%%%%%%%%%%%%%%%%%%
%%%%%%%%%%%%%%%%%%%%%%%%%%%%%%%%%%%%%%%%%%%%%%%%%%%%%%%%%%%%%%%%%%%%%%%%%%%%%%

\section{Results and discussions}

\subsection{Insulating ground states}
The ground state of the system defined by the Hamiltonian in Eq. (\ref{ham}),
for the case of $J/U\approx 0$, can be analytically identified using the method
of strong coupling perturbative expansion. As discussed earlier the system
supports two possible solid phases the CDW and SDW. The phase with the 
lower energy determines the ground state of the system. Consider the case of
$n_A=1$ and $n_B=0$, and considering that $V_1/V_2=2\sqrt{2}$ the energies of 
the two phases given by Eqs. (\ref{energy_checkerboard}) and 
(\ref{energy_striped}) are simplified to

\begin{eqnarray}
             E_{\text{CDW}}&=&-8\mu+5.65V_1-20.17\frac{J^{2}}{V_{1}},
             \label{energy_checkerboard_reduced}  \\
             E_{\text{SDW}}&=&-8\mu+8V_{1} +24.75\frac{J^{2}}{V_1} 
             +\frac{32J^{2}}{-1+V_{1}}.
             \label{energy_striped_reduced}
\end{eqnarray}

To compare the energies of the two states given above, we first compare the
zeroth order energies.  We express the long-range interaction terms in terms 
of $V_{1}$ in both the expressions and we find that 
$\frac{8}{\sqrt{2}}V_{1}<8V_{1}$. This shows that zeroth order energy of 
(\ref{energy_checkerboard_reduced})  is lower than zeroth order energy of
(\ref{energy_striped_reduced}). Thus, at the zeroth order, the CDW is the 
favoured state. Next, let us compare the second order energy correction
between the two expressions after scaling all the long-range interaction terms 
with $V_{1}$. In this case too, the CDW has lower energy. The trend of the 
energies corresponding to the CDW and SDW phases obtained from 
Eqs. (\ref{energy_checkerboard_reduced}) and (\ref{energy_striped_reduced})
are shown in Fig. \ref{ener_compare}. From the figure it is evident that 
CDW has lower energy for all values of $J/U$. Thus, with the dipolar form
of isotropic interaction CDW is the favoured ground state. This is along the 
expected lines as the NN interaction favours CDW, and with the dipolar form 
of long-range interaction the NNN interaction cannot lower the energy of the
SDW to emerge as the ground state.
\begin{figure}[h]
   \begin{center}
      \includegraphics[width=8.5cm]{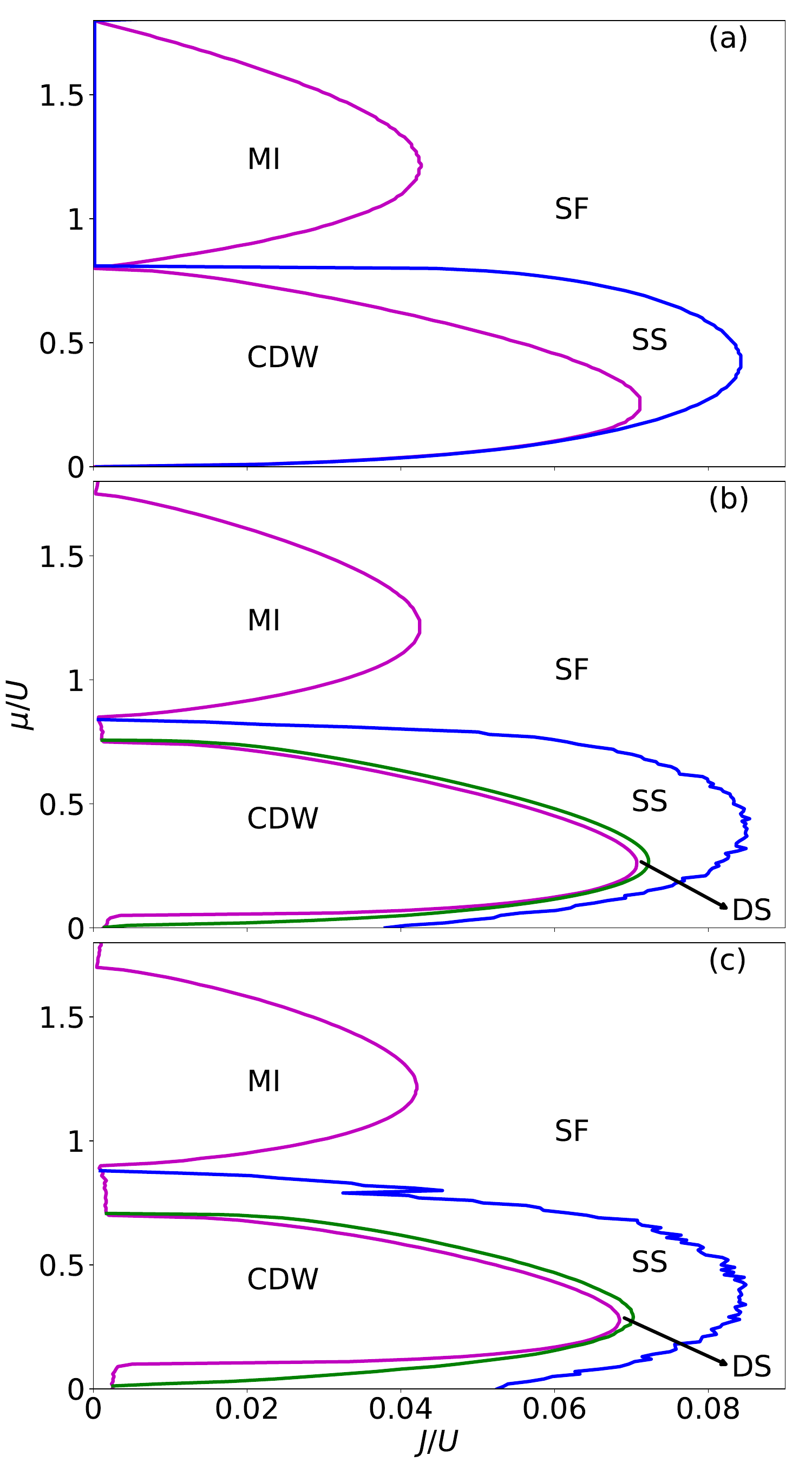}
      \caption{Ground state phase diagram of DEBHM for $V_{1}/U=0.2$ and 
               $D/U = 0$ (a), $0.05$ (b) and $0.1$ (c). With non-zero disorder,
               the DS phase appears in the phase diagram, forming an envelope 
               around the CDW phase.}
               \label{phase-v1-d}
	\end{center}
\end{figure}

%%%%%%%%%%%%%%%%%%%%%%%%%%%%%%%%%%%%%%%%%%%%%%%%%%%%%%%%%%%%%%%%%%%%%%%%%%%%%%
%%%%%%%%%%          Results and discussions                %%%%%%%%%%%%%%%%%%%
%%%%%%%%%%%%%%%%%%%%%%%%%%%%%%%%%%%%%%%%%%%%%%%%%%%%%%%%%%%%%%%%%%%%%%%%%%%%%%
\subsection{Phase diagram from SGMF}
To calculate the ground state with SGMF, the parameters in the DEBHM 
Hamiltonian are scaled with respect to the on-site interaction energy $U$. 
Thus the relevant parameters are $J/U$, $\mu/U$, $D/U$, $V_{1}/U$ and 
$V_{2}/U$. For the present work, we consider a system size of $50 \times 50$.
This implies that the system consists of 2500 lattice sites and this is large
enough to provide reliable statistics. To introduce disorder, each lattice 
site is assigned a random number from a univariate distribution within the 
bound $[-D,D]$. As mentioned earlier, the disorder combines with the chemical 
potential and yields the effective chemical potential of a lattice site.
One difficulty associated with disordered systems is, one has to do a 
disorder averaging by repeating the calculations with different disorder
realizations. To obtain reliable average properties the averaging must be 
done over a large number of samples. This requirement is to be balanced with 
the computational cost of considering a large system size. For the mean-field
calculation the dimension of the Fock space is $N_{b}=5$. And, the disorder
averaging of the order parameters and properties are done over 50 disorder 
samples.
\begin{figure}[h]
  \includegraphics[width=8.5 cm]{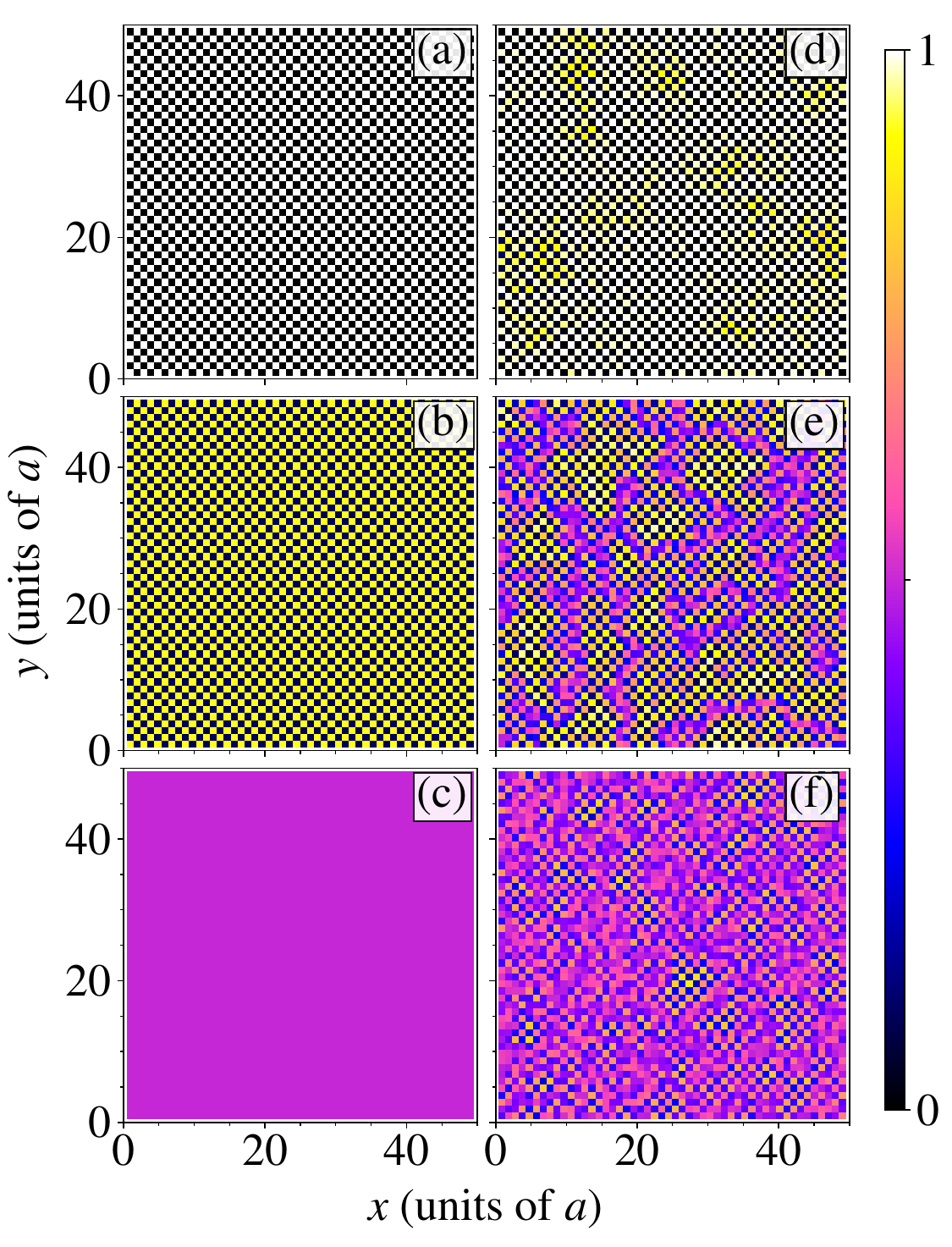}
   \caption{Distribution of bosons on a 50 $\times$ 50 
         lattice for $\mu/U$=0.29 and $V_{1}/U$=0.2. Panels (a), (b) and (c) 
         shows the CDW, SS and SF phases at $J/U= 0.07, 0.075$ and $0.087$ 
         respectively, with $D/U=0$. Panels (d), (e) and (f) shows the DS, SS
         and SF phases with the disorder strength $D/U=0.1$ for 
         $J/U=0.07,0.075$ and $0.087$ respectively.}
   \label{densityplot_lrd}
\end{figure}

After identifying the ground state quantum phases, we generate the phase
diagram in the $J/U-\mu/U$-plane by selecting a suitable order parameter to 
demarcate the phase boundaries. Using the order parameter, for a specific 
value of $\mu/U$, we employ bisection method along $J/U$ to locate the critical 
point between two neighbouring quantum phases. In the case of 
incompressible-compressible phase boundary we use the system size averaged 
SF order parameter $\phi$ as the order parameter and threshold of 
$\phi \leqslant 10^{-4}$ in the bisection method. In the phase
diagram, the DW lobes are enveloped by DS, and the DW-DS transition is of
incompressible-compressible type so $\phi$ can be used to chart the
phase boundary. Similarly, for the MI phase it transitions to BG, which
is a compressible phase. The DS phase, in the phase diagram, is followed by 
the SS phase, and we choose $|S(\pi,\pi)|$ as the order parameter to 
distinguish these two phases. As the DS phase is a solid phase with rare SF 
islands, it has higher density contrast and hence, possesses larger value of
$|S(\pi,\pi)|$ than SS. Consequently, we set $|S(\pi,\pi)| \geqslant 0.9$ as 
the threshold in the bisection method to determine the DS-SS phase boundary.
On increasing $J/U$ further, we encounter the SS-SF phase boundary 
and for the determination of 
phase boundary between SS and SF, we 
choose $\left| S(\pi,\pi) \right|$ as the order parameter and we consider 
$| S(\pi,\pi)| \geq 10^{-3}$ as the threshold to determine the phase boundary 
between SS and SF phase.

To determine the effects of long-range interactions we begin by examining
the quantum phases of DBHM, in which only the on-site interactions is
considered. The phase diagram of the system for $D/U = 0.2$ 
is shown in Fig.~\ref{phase-dbhm}. It shows the existence of 
BG phase for low $J \lesssim 0.005U$ in a small region between the 
incompressible Mott lobes and it also exist as a thin patina surrounding
the Mott lobes. The latter is consistent with the theorem of inclusions
\cite{pollet_09}. 
According to which in presence of disorder the MI-SF transition 
is intervened by the BG phase. As expected, the DW phase is 
absent without the long-range interactions and consequently, 
the DS phase does not occur in the phase diagram. The 
quantum phases with diagonal long-range order emerge in the 
system with the introduction of the long-range interactions, and these
are examined in detail.
\begin{figure}[h] 
   \includegraphics[width=8.5cm]{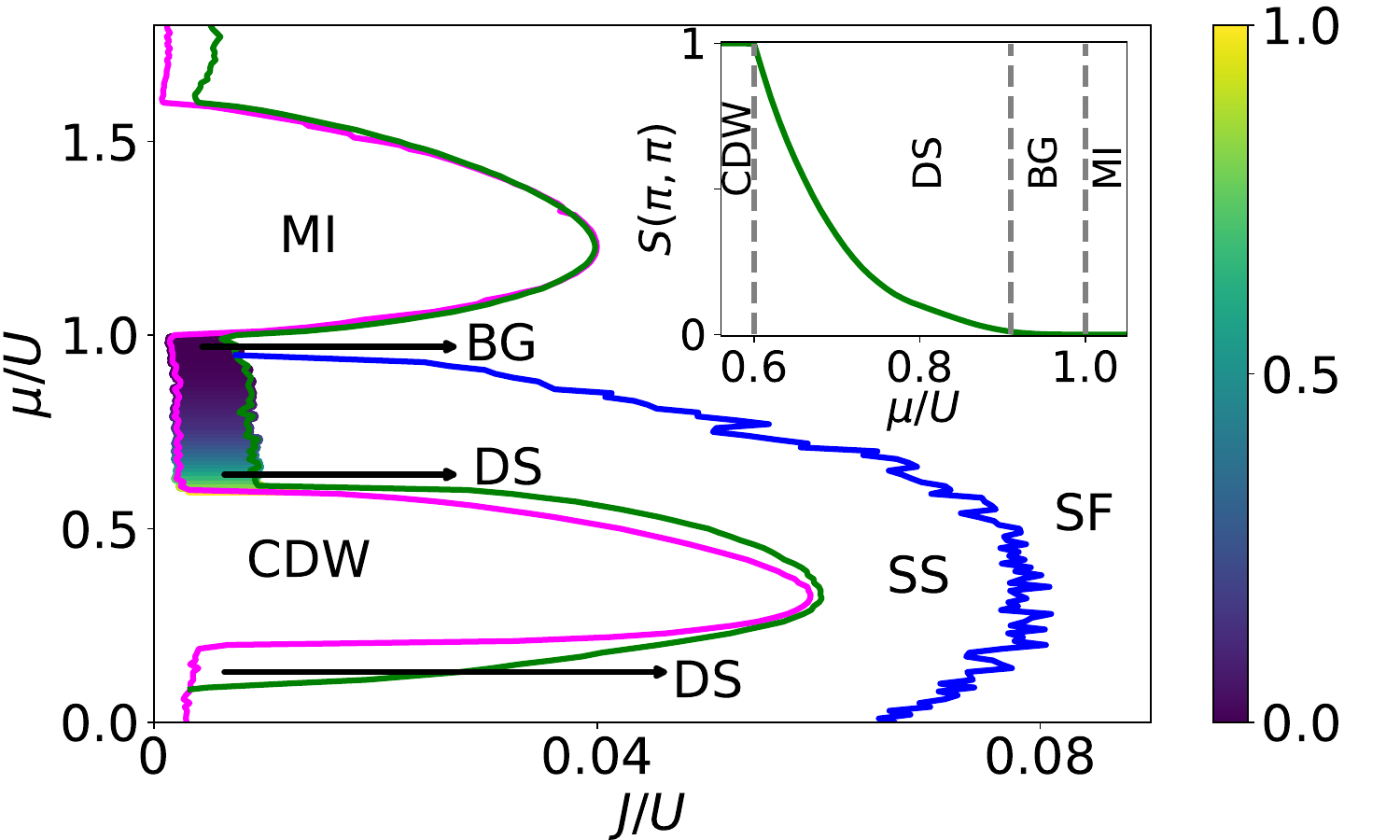}
    \caption{Ground state phase diagram of DEBHM for $V_{1}/U=0.2$ and 
    $D/U=0.2$.The violet phase boundary  
    indicates the phase boundary between the insulating phases (CDW,MI) 
    and disordered phases (DS,BG). The green phase boundary indicates the  
    phase boundary between disordered phases and supersolid and superfluid.  
    The blue phase boundary indicates the phase boundary between SS and SF.  
    The color gradient between the CDW lobe and MI lobe is plotted based on  
    $\left|S(\pi,\pi)\right|$ and it indicates the transition from DS to BG  
    in between the lobes. The inset shows the plot of the        
    structure factor $S(\pi,\pi)$ at $J/U=0.005$ as a function of 
    the $\mu/U$.}
    \label{phase-ds-bg}
\end{figure}

%%%%%%%%%%%%%%%%%%%%%%%%%%%%%%%%%%%%%%%%%%%%%%%%%%%%%%%%%%%%%%%%%%%%%%%%%%%%%%

%%%%             Subsubsection: Nearest neighbour interaction             %%%%
%%%%%%%%%%%%%%%%%%%%%%%%%%%%%%%%%%%%%%%%%%%%%%%%%%%%%%%%%%%%%%%%%%%%%%%%%%%%%%

\subsubsection{Nearest neighbour interaction}
The distinguishing feature of DEBHM is the emergence of the DS phase 
characterized by zero $\rho_{s}$, but finite $\phi_{p,q}$ and $S(\pi,\pi)$. 
Using SGMF we first generate ground state phase diagrams of the system by 
considering NN interaction, that is, only $V_{1}$ is non-zero. For $zV_{1}<1$,
where $z$ is the co-ordination number, the model has CDW phase or MI phase 
as the insulating quantum phase of the ground state. In the absence of 
disorder ($D/U = 0$) the CDW phase undergoes a quantum phase transition to 
the SS phase when $J/U$ is increased. The transition is evident from the 
phase diagram shown in Fig.\ref{phase-v1-d}(a). However, as discernible from
the phase diagram in Fig.\ref{phase-v1-d}(b), the introduction of disorder
with $D/U=0.05$ leads to the emergence of the DS phase and it intervens the 
transition from CDW to SS. This is akin to BG phase intervening the transition 
from MI to SF when disorder is introduced to BHM. The DS phase is compressible 
but possesses the character of a solid. The latter is associated with diagonal
long-range order and characterized by finite structure factor. The solid 
lobes shrink further as the strength of the disorder is increased, which is
evident from the phase diagram for $D/U=0.1$ as shown in 
Fig.\ref{phase-v1-d}(c). However, as noticeable in the figure, the SS phase 
is enhanced. For illustration, the density distribution for some of the 
quantum phases, present in the phase diagrams shown in Fig.\ref{phase-v1-d}, 
are shown in Fig.\ref{densityplot_lrd}.

\begin{figure*}[htbp] % 'figure*' to span across two columns
    \centering
    \includegraphics[width=17 cm]{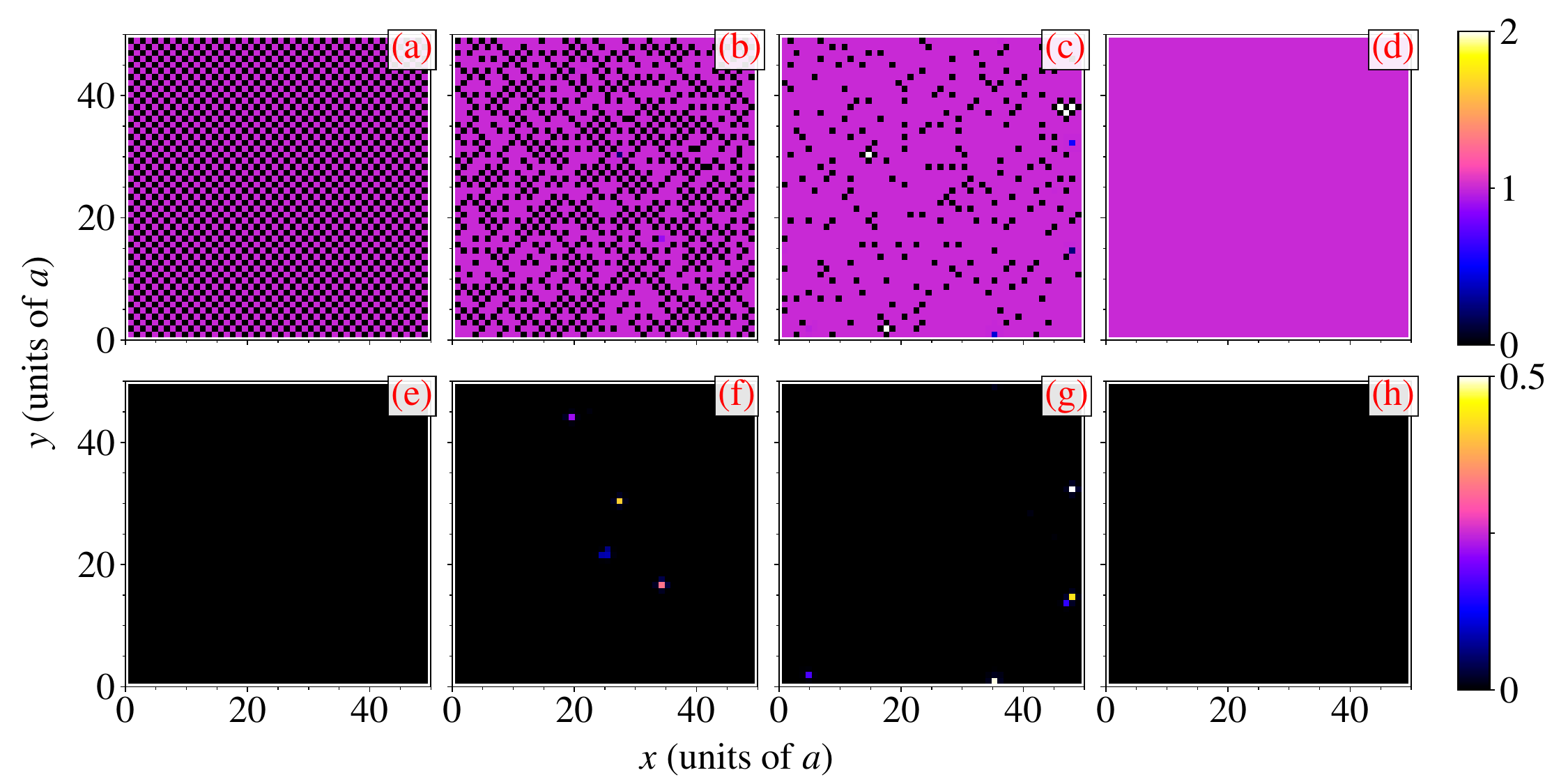}
     \caption{Boson density and order parameter
        plots at $J/U=0.005$,$D/U=0.2$ and $V_{1}/U=0.2$.
	Panels (a)-(d) shows the density plots and (e)-(h) shows the 
	SF order parameter plots
	for $\mu/U$ = 0.55, 0.75, 0.96 and 1.04 
	corresponding to the CDW, DS, BG and MI phases respectively.}
    \label{ds_bg_density_phi_plots}
\end{figure*}

The presence of the BG phase around the MI lobe is not obvious for $D/U=0.1$. 
However, as the disorder is increased the existence of the BG phase becomes 
more prominent. Thus, as shown in Fig.\ref{phase-ds-bg}, the BG phase 
surrounding the MI phase is noticeable for $D/U=0.2$. And, similarly, the 
domain of the DS phase around the 
CDW lobe is larger compared to $D/U=0.1$.  As mentioned earlier, $\phi$ 
is used to differentiate the compressible BG phase from the incompressible MI phase. 
And, the superfluid fraction $|f_{s}|$ given in Eq.(\ref{sf-stiffness}) is 
used as the order parameter to segregate the BG phase from other compressible 
phases like SF and SS. We choose $|f_{s}| \leqslant 10^{-7}$ as the threshold 
to determine the phase boundary between BG and other compressible phases. 
In contrast to the compressible phases, the incompressible lobes
of CDW and MI phases shrinks and in between these lobes there is a strip across
which DS-BG transition occurs. The transition is indicated by the change
in  $|S(\pi,\pi)|$ as we trace it across the strip from 
lower $\mu/U$ to higher value for fixed $J/U$ as shown in the inset of 
Fig.\ref{phase-ds-bg}. As seen from the figure, within CDW $|S(\pi,\pi)|$ is 
unity and decreases rapidly across the DS, and it is zero in the BG and 
MI phases. The value of $| S(\pi,\pi)|$ is zero within BG as it has no 
diagonal long range order. Thus $|S(\pi,\pi)|$ becoming zero signals the 
transition from DS to BG phase. In the MI phase $|S(\pi,\pi)|$ is zero as it
is a uniform density phase. The density and SF order parameter distributions 
of the quantum phases in this strip are shown in 
Fig.~\ref{ds_bg_density_phi_plots}.

\begin{figure}[h]             
     \begin{center}
     \includegraphics[width=8.5cm]{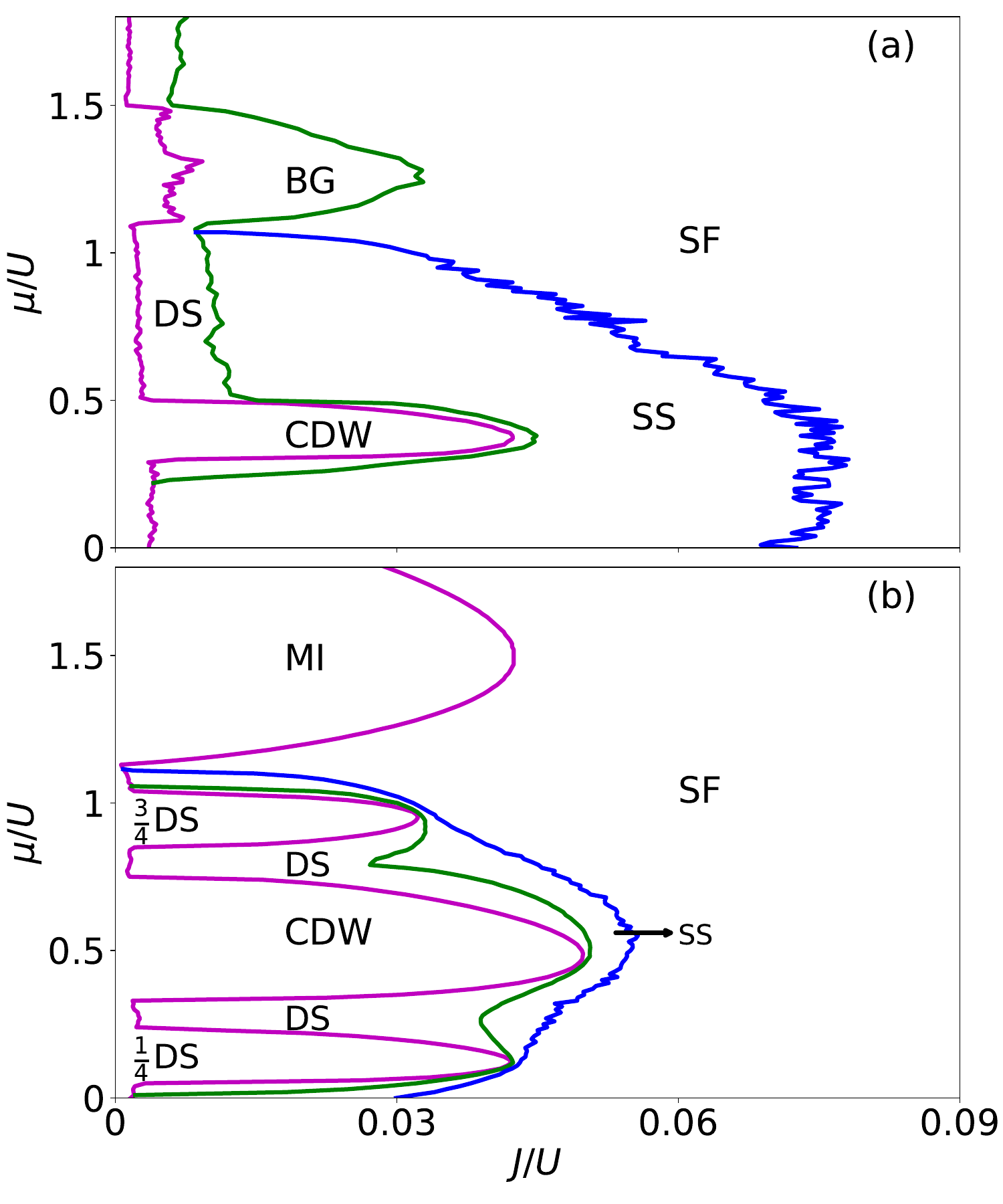}       
	     \caption{Panel (a) shows ground state phase diagram 
	     of DEBHM for $V_{1}/U=0.2$ and $D/U=0.3$ and (b) shows
	     the phase diagram at $V_{1}/U=0.2$, $V_{2}/U=0.0707$ and
	     $D/U=0.05$.}
      \label{phase-high-d-phase-nnn}                                           
     \end{center}                                                               
\end{figure}

To study the effect of disorder on various quantum phases, we increase the 
disorder strength to $D/U=0.3$, while retaining $V_1=0.2U$. The resulting 
phase diagram is shown in Fig.(\ref{phase-high-d-phase-nnn})(a). The
figure shows that MI phase shrinks significantly and survives only in a small 
region. Notably, the MI phase at larger $J/U$ is replaced by the BG phase, 
thereby expanding the BG domain surrounding the MI. The CDW lobe also 
diminishes, although it continues to occupy a relatively larger area, as its 
region between the solid lobes expands. In contrast, the SS phase becomes 
more prominent, now covering a larger portion of the phase diagram. Thus the 
disorder has a larger impact on the MI phase than the CDW phase. And, the SS 
phase with both diagonal and off-diagonal long-range orders is favored. The  
sensitivity of MI phase to disorder is due to the ease with which the 
commensurate filling can be disrupted. In comparison, The CDW phase 
maintains the diagonal long-range order and incompressibility even under
increasing disorder, as its destruction requires rearrangement of particle
occupancies across a larger number of lattice sites.

%%%%%%%%%%%%%%%%%%%%%%%%%%%%%%%%%%%%%%%%%%%%%%%%%%%%%%%%%%%%%%%%%%%%%%%%%%%%%%
%%%%  Subsubsection: Nearest and next-nearest neighbour interactions      %%%%
%%%%%%%%%%%%%%%%%%%%%%%%%%%%%%%%%%%%%%%%%%%%%%%%%%%%%%%%%%%%%%%%%%%%%%%%%%%%%%
\subsubsection{Nearest and next-nearest neighbour interaction}
To examine the impact of the long-range interactions further, we introduce 
NNN interaction by considering $V_{2}= V_{1}/(2\sqrt{2})$. This is the
relation when the long-range interaction is of dipole-dipole type having
$1/r^3$ radial dependence. The DEBHM then exhibits four types of insulating 
quantum phases, these are $1/4$ DW, CDW, $3/4$ DW and MI. The phase diagram
for the case of $V_1/U=0.2$, which implies $V_2/U= 0.0707$, and disorder
strength $D/U=0.05$ is shown in Fig.\ref{phase-high-d-phase-nnn} (b). 
In this case, $|S(\pi,\pi)| \geqslant 0.45$ is considered as the threshold to 
determine the phase boundary between DS and SS. From the figure, it is evident 
that with NNN interaction the SS lobe shrinks. Hence, we can infer that 
increasing the range of the long-range interaction at fixed disorder strength, 
promotes SF phase and suppresses the SS phase.

%%%%%%%%%%%%%%%%%%%%%%%%%%%%%%%%%%%%%%%%%%%%%%%%%%%%%%%%%%%%%%%%%%%%%%%%%%%%%%
%%%%             Subsubsection: Nearest neighbour interaction             %%%%
%%%%%%%%%%%%%%%%%%%%%%%%%%%%%%%%%%%%%%%%%%%%%%%%%%%%%%%%%%%%%%%%%%%%%%%%%%%%%%

\subsection{ DS-BG transition}

\begin{figure}[h]
    \begin{center}
       \includegraphics[width=8.5cm]{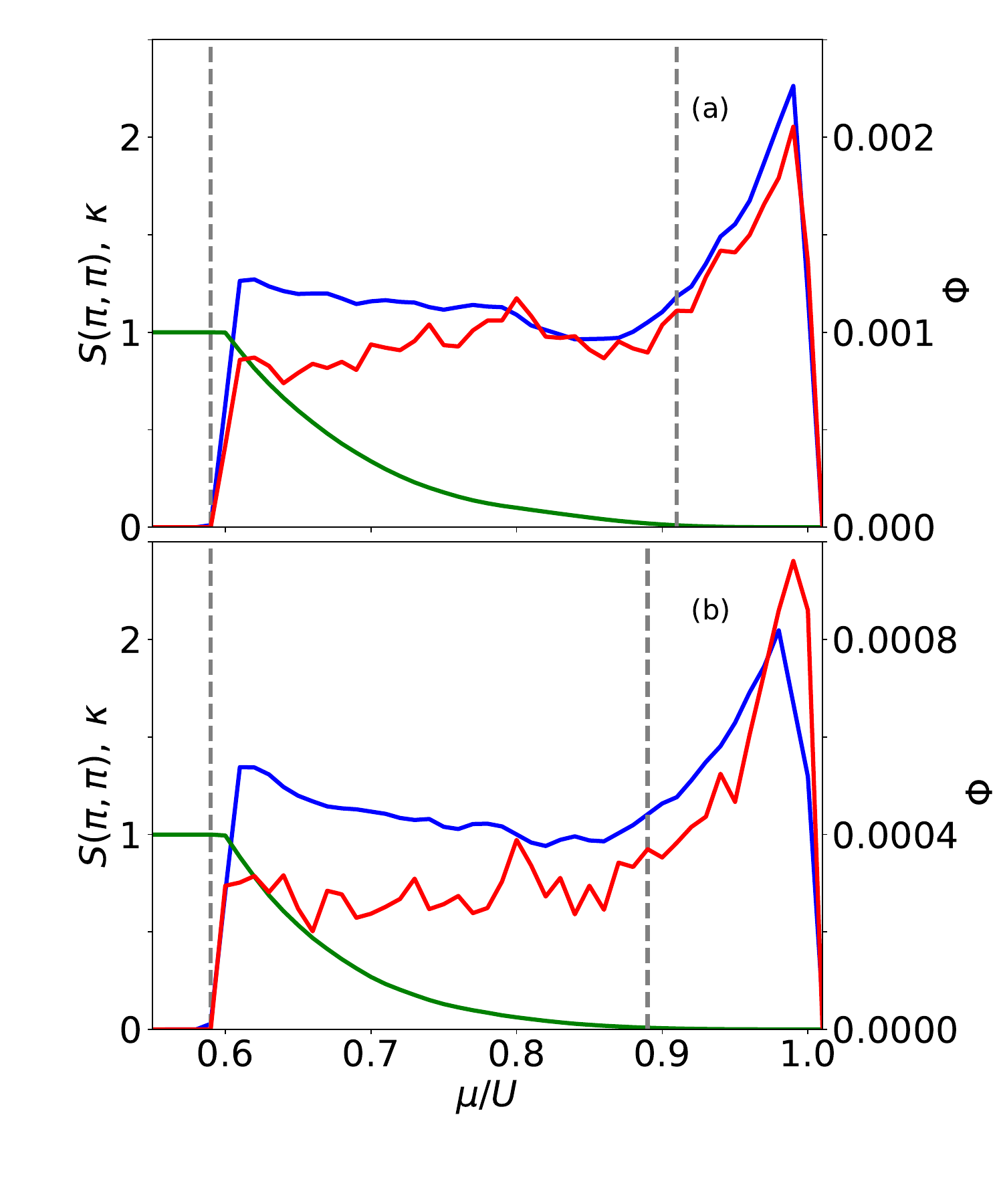}
       \caption{ Panel (a) and (b) displays the disorder averaged 
	       properties of the system based on SGMF and CGMF results 
               respectively for $V_{1}=0.2U$ and $D=0.2U$. Green, blue and 
               red curves represent $S(\pi,\pi)$, $\kappa$ and $\Phi$.}
       \label{compress_vs_mu_plot}
    \end{center}                                                               
\end{figure}                    

To analyse the DS-BG transition which occurs across a vertical strip 
at $\mu=0.91U$ in better detail we examine and compare the trends of 
various properties. As discussed in the previous section, the DS-BG 
transition is evident from the phase diagram shown in Fig. \ref{phase-ds-bg}. 
One of the important properties is the compressibility $\kappa$, and its 
variation from DS to BG phase for a fixed value of $J/U$ is shown by the blue 
curve in Fig. \ref{compress_vs_mu_plot}(a). In the figure, across the DS domain
$\kappa$ decreases reaching a minima and then, increases sharply on entering 
the BG domain. This indicates, hardening of the DS phase as it evolves 
towards the BG phase and reaches a minima before softening.
In the DS phase, $\kappa$ is finite as the disorder modifies $\mu$ to the
effective chemical potential $\tilde{\mu}$ and which may correspond to the
SF or MI phase of the clean eBHM. With the modified $\mu$, some of the vacant 
lattice sites in the CDW can then acquire finite occupancies. The 
disorder averaged SF order parameter $\Phi$ is non-zero if 
some of the vacant lattices acquire 
fractional occupancies and along the strip the trend  of $\Phi$ is shown by the
red curve in Fig.~\ref{compress_vs_mu_plot}(a). Additional vacant sites are filled 
at higher $\mu$ and reduces the probability of finding a vacant site. Within
the DS phase the occupancy is bounded to unity upto a higher value 
of $\mu$ and this limits the possiblity to increase the number of particles.
As a result, the increase in number of particles in the system slows
down with increasing $\mu$ and $\kappa$ is expected to decrease. This trend 
continues till the system enters BG phase at $\mu \approx 0.91U$, where few 
sites with large $\tilde{\mu}$ host more than unit occupancy leading to a 
sharp increase in $\kappa$ and $\Phi$.

To check the robustness of the DS-BG phase transition, we examine the 
transition with the CGMF method.  The method incorporates inter-site 
correlations within a cluster exactly and thus, can provide better description
of the phase transition than the SGMF method. The discussions on CGMF method
and relevant details, in particular, the numerical considerations are 
given in previous work~\cite{luhmann_13,bai_18}. The CGMF results based on  
cluster size of $2\times2$ suggests that both the DS and BG phases are
robust and survives the beyond mean-field effects. With this cluster size and 
$N_B =4$, the properties of the system as it undergoes DS-BG transition is 
shown in Fig.~\ref{compress_vs_mu_plot}(b). The trend in $S(\pi,\pi)$ suggests 
a transition from the structured DS phase to BG phase at $\mu\approx 0.89U$, 
which is lower than the corresponding SGMF result. The trend in $\kappa$ and 
$\Phi$ also shows, like in the SGMF case, a sharp increase in the BG phase.

\begin{figure}[H]
    \begin{center}
     \includegraphics[width=8.5cm]{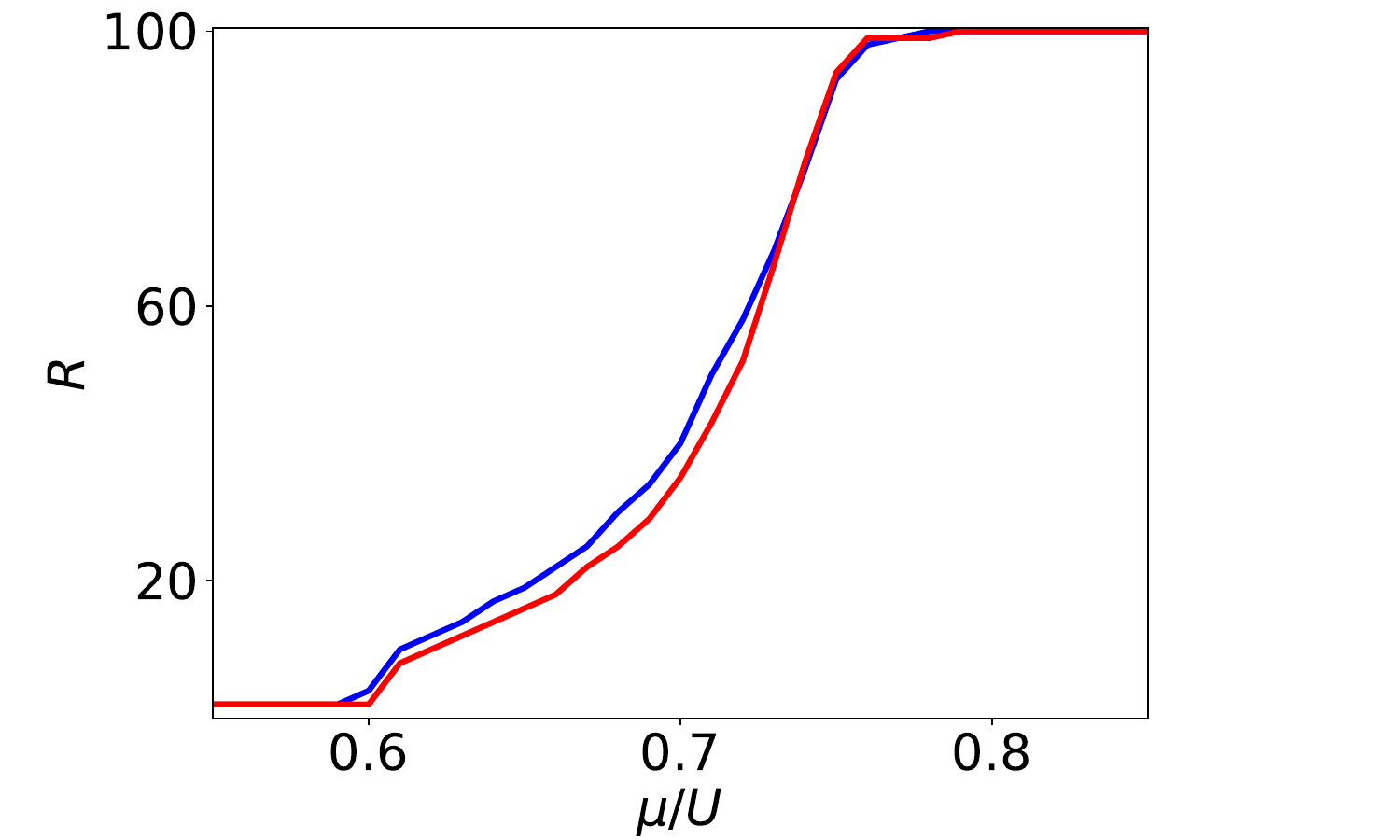}
      \caption{DS-BG transition as a percolation transition.
               The span $R$ of the largest unit occupancy domain is plotted
	       as a function of $\mu$ for $V_{1}=0.2$ and $D=0.2$. For a 
	       $50 \times 50$ lattice, spanning domain along the two 
	       directions would correspond to $R=100$ and this occurs at
	       $\mu = 0.8U$. The blue (red) curve are computed using the 
	       CGMF (SGMF) results.}
      \label{percol}
    \end{center}
\end{figure}

The trasition from DS to BG is an example of a phase transition which 
destroys diagonal long-range order and can be studied using tools from 
percolation theory. In the DS phase, rare 
island of superfluidity forms on a background of the checkerboard solid. This
structured backround is eventually destroyed as the system enters the BG phase
where the rare SF islands are formed over a MI background. In the BG phase,
the MI background comprising of unit occupancy can be seen as a single domain 
which spans the entire lattice, while in DS solid phase, the domains of unit
occupancies are small as evident in Fig.~\ref{ds_bg_density_phi_plots}(b,c). 
Thus, the transition of the background phase from checkerboard to MI can be 
viewed as a percolation of the unit occupancy domains and the span ($R$) of 
the largest domain, along the two directions of the lattice would be a 
property of interest. To identify the domains of unit occupancy and for 
analysing the span of the largest domains, we use the algorithm discussed in 
ref.~\cite{sable_23}. In Fig.~\ref{percol}, we plot disorder averaged span 
of the largest domain as a function of $\mu$. The figure shows a percolation 
transition occuring at $\mu\approx 0.8U$, which is lower than the DS-BG 
transition point determined using structure factor with a cut-off 
$S(\pi,\pi) = 0.01$. This is as expected, the percolation analysis determines
the onset of the transition from checkerboard to MI phase of the background
phase. The percolation transition indicates that there is a large enough 
domain of MI which extend the entire system along both directions. 

%%%%%%%%%%%%%%%%%%%%%%%%%%%%%%%%%%%%%%%%%%%%%%%%%%%%%%%%%%%%%%%%%%%%%%%%%%%%%%
%%%%                   section: Conclusion                                %%%%
%%%%%%%%%%%%%%%%%%%%%%%%%%%%%%%%%%%%%%%%%%%%%%%%%%%%%%%%%%%%%%%%%%%%%%%%%%%%%%

\section{Conclusions}
We have examined the zero-temperature quantum phases and quantum phase 
transitions of soft-core dipolar bosons in a 2D disordered square optical
lattice using DEBHM. We determine the lower energy configuration of bosons 
in the incompressible or solid phase analytically using the strong-coupling 
perturbative expansion. We find that the CDW phase is favoured. 
With the long-range dipolar interactions two types of disordered
phases DS and BG emerge. Among these the DS phase has diagonal long-range 
order. The DS and BG phase emerges when the effective chemical potential 
(resulting from disorder in lattice potential) allow few of the lattice 
sites to support superfluidity. At higher disorder $D/U=0.2$, there is a strip
adjacent to $J/U=0$ along which DS-BG occurs. The robustness of these phases 
and DS-BG transition is verified using the CGMF method. To understand the
DS-BG transition in better detail we have examined various properties of 
the system using both SGMF and CGMF. The structured CB background phase of the
DS undergoes a transition to the homogeneous MI as the system enters BG phase.
Thus, the transition of the background phase can be analysed from the
perspective of percolation of system by domains of unit occupancy. The 
percolation results underestimate the transition point, which suggests that 
the transition should occur after the spanning cluster grows to become
the uniform background seen in the MI phase. We also studied the fate
of quantum phases at high disorder $D/U = 0.3$ and observe that the 
MI lobe shrinks and it is replaced by the BG phase. The CDW lobe, on the other
hand, retains the extant of the domain. This indicates that CDW phase is more 
robust against disorder.

\bibliography{ref}{}

\end{document}